# Simultaneous multiplane imaging with reverberation multiphoton microscopy

Devin R. Beaulieu[1]*, Ian G. Davison[2,3], Thomas G. Bifano[3], and Jerome Mertz[4,3]

[1]Dept. of Electrical Engineering, Boston University, 8 St. Mary's St., Boston MA 02215
[2]Dept. of Biology, Boston University, 5 Cummington Mall, Boston MA 02215
[3]Photonics Center, 8 St. Mary's St. Boston MA 02215
[4]Dept. of Biomedical Engineering, Boston University, 44 Cummington Mall, Boston MA 02215
*Corresponding author: drbeau@bu.edu

**Multiphoton microscopy (MPM) has gained enormous popularity over the years for its capacity to provide high resolution images from deep within scattering samples[1]. However, MPM is generally based on single-point laser-focus scanning, which is intrinsically slow. While imaging speeds as fast as video rate have become routine for 2D planar imaging, such speeds have so far been unattainable for 3D volumetric imaging without severely compromising microscope performance[2]. We demonstrate here 3D volumetric (multiplane) imaging at the same speed as 2D planar (single plane) imaging, with minimal compromise in performance. Specifically, multiple planes are acquired by near-instantaneous axial scanning while maintaining 3D micron-scale resolution. Our technique, called reverberation MPM, is well adapted for large-scale imaging in scattering media with low repetition-rate lasers, and can be implemented with conventional MPM as a simple add-on.**

The standard method for obtaining volumetric images with MPM is to perform x-y scanning with galvanometric mirrors, and then z-scanning by adjusting the microscope objective; this is slow and cumbersome. Faster volumetric imaging can be obtained by purposefully decreasing image resolution[3], or by using faster z-scanning mechanisms, such as electrically tunable lenses[4], deformable mirrors[5], voice-coils[6], or tunable acoustic gradient (TAG) lenses[7]. For example, although TAG can provide axial scan rates at tens of kilohertz, it comes at the cost of limited depth range[8].

Alternatively, simultaneous multifocus[9], extended focus[10,11], or stereoscopic[12,13] illumination can be achieved by wavefront engineering, providing 2D images of volumetric samples obtained from single transverse scans. While fast, these solutions sacrifice axial resolution by yielding only 2D projections. Axial localization and segmentation can be calculated post acquisition, but with the requirement of computational models and/or a priori knowledge about the sample structure. Consequently, such solutions involving simultaneous multiplexed illumination are best suited for sparse samples.

The use of high-speed detection electronics has opened new approaches for near-simultaneous multiplexing, taking advantage of the ability to individually measure fluorescence signals a few nanoseconds apart. This has been implemented in previous work by separating the illumination beam into a few (usually two) beamlets of different pathlengths[14,15,16,17,18]. In this manner, the signals produced by each beamlet can be separated in time using fast detection electronics. By focusing each beamlet to a different depth within the sample, a near simultaneous focal stack can be obtained from a single transverse scan. However, such multiplexing becomes technically cumbersome with increasing number of beamlets, and leads to laser power loss when the number of beamlets is greater than two (unless the focal planes are staggered in the transverse direction[16]). A similar multiplexing approach has been implemented in the detection optics of a camera-based imaging system[19].

We present here a simplified alternative to the above temporal multiplexing solutions that makes use of a reverberation loop. This more general approach provides an infinite series of beam foci, performing a near-instantaneous axial scan, while delivering the full illumination power to the sample. A diagram of our system is shown in Figure 1. The optical configuration is typical of MPM setups except for the addition of a reverberation loop upstream from the beam steering mirrors. Here a 50:50 non-polarizing beamsplitter splits the illumination beam, with half the light proceeding to the sample normally, and the other half entering the loop. The 1× relays in the loop are intentionally mis-adjusted (spaced too far apart) so that a small amount of focus is added to the beam. Upon returning to the beamsplitter half of the light exits the loop and proceeds to the sample, but now with a modified focus and time delay (due to the time spent in the loop). The light remaining in the loop continuously repeats the process, accumulating slightly more focus and delay upon each pass.

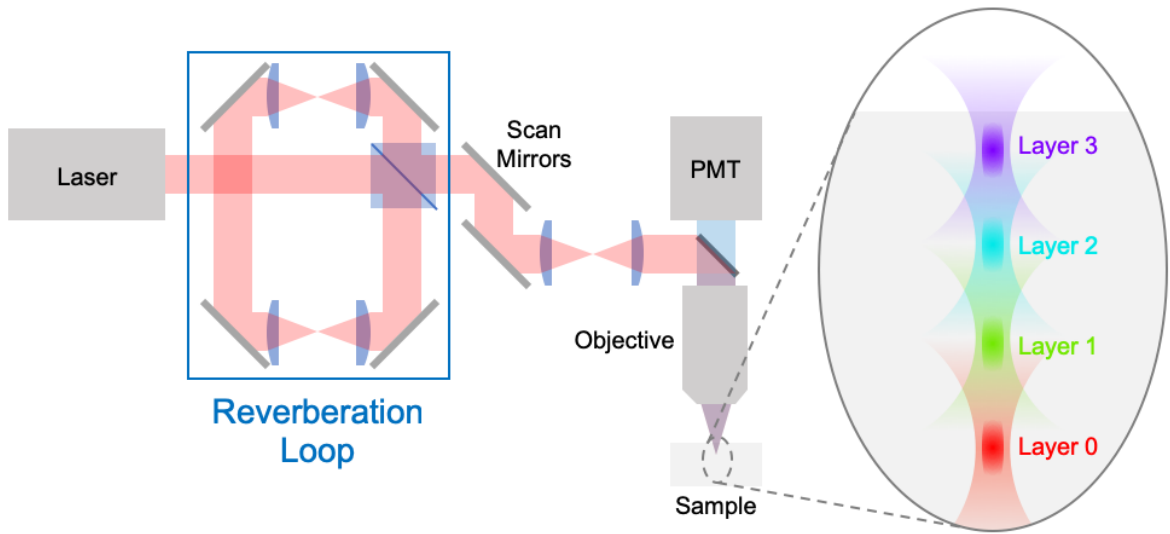

Figure 1: Reverberation MPM schematic. For each laser pulse, the reverberation loop creates an infinite series of beam foci separated in space and time (only four are shown here, terminated by the sample surface). The spatial separation $\Delta z$ between each focus in the sample can be controlled as desired without affecting alignment, by adjusting the pathlength of the loop (the left pair of mirrors and lenses in the loop are mounted on a linear translation stage – see Supplementary Information).

As a result of the reverberation loop, each laser pulse produces a series of beam foci of decreasing depth within the sample that arrive sequentially in time. The incident power associated with the $n$-th focal spot is given by $P_n = 2^{-(n+1)} P_{in}$, where $P_{in}$ is the total laser power incident on the loop and $n = 0$ corresponds to the deepest layer in the sample. In MPM, only the ballistic (i.e. unscattered) portion of this power contributes to fluorescence generation[1]. The relative fluorescence power produced at each focal spot is thus given by $F_n = F_0 \exp[mn(\Delta z/l_s - \ln(2))]$, where $m$ is the nonlinear order ($m = 2$ for two-photon microscopy), $l_s$ is the scattering mean-free-path at the illumination wavelength, and we have assumed a roughly homogeneous fluorescence labeling density. In other words, even though the incident power associated with each focal spot decreases geometrically with decreasing depth (increasing $n$), the resulting fluorescence may or may not decrease depending on our choice of $\Delta z$. For example, if the inter-layer spacing is chosen such that $\Delta z = l_s \ln 2$, the decrease in scattering at shallower depths exactly compensates for the decrease in incident power with increasing $n$, and the fluorescence produced from each focal spot remains roughly constant at all depths. On the other hand, if a finer inter-layer spacing is desired (i.e. $\Delta z < l_s \ln 2$), the fluorescence becomes successively dimmer with shallower depths, which can be corrected in post processing provided the detector supplies adequate dynamic range (see Supplementary Information).

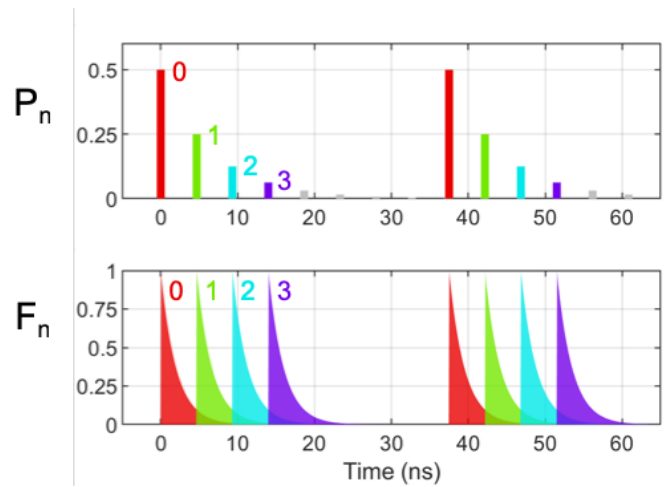

Figure 2: Timing and power of illumination pulses and corresponding fluorescence signals (assuming, for example, a 4 ns fluorescence lifetime). Here, $\Delta z$ is chosen so that the reduction in illumination power between layers is exactly offset by the reduction in scattering from shallower foci. Note that shorter delays allow more layers to fit between each laser pulse, but at the cost of additional crosstalk from previous layers.

Figure 2 illustrates the timing of the illumination and fluorescence pulses, with a different color indicating each focal depth (or layer). The 1.4 m long reverberation loop used for this experiment produces a 4.7 ns delay between focal depths, allowing each to be measured individually using a high-speed amplifier (1 GHz bandwidth) and digitizer (1.5 GS/s). In theory, the pulse reverberation subsists indefinitely (with decreasing power), producing an arbitrary number of focal depths. In practice, the sequence of focal depths is terminated at $n$ when the $(n + 1)$-th focal spot exits the sample, thus terminating the sequence of fluorescence and preventing it from overlapping with signal from the next laser pulse (alternatively, if $\Delta z < l_s \ln 2$, the fluorescence can fade away before such overlap occurs).

But adequately fitting the number of focal spots between laser pulses is not the only constraint. We must also bear in mind that the fluorescence lifetime of fluorescent indicators is typically a few nanoseconds[20]. To properly distinguish the signal from successive focal spots, the time delay between these should be longer than the fluorescence lifetime. In our setup, the fluorescence signal was integrated over time bins of durations up to 4.7 ns, corresponding to the reverberation delay, allowing us capture most of the fluorescence produced by each focal spot while maintaining a small crosstalk between successive spots of typically less than 9%. We note that most of this crosstalk can be removed in post-processing, by subtracting a proportion of the previous layer from each layer (see Supplementary Information).

The dual constraints of maximizing number of layers between laser pulses while minimizing inter-layer fluorescence crosstalk motivate the use of lasers with slower repetition rates and correspondingly higher pulse powers. As it happens, such lasers are advantageous for deep imaging[21,22], and even indispensable for three-photon imaging[23]. To achieve lower repetition rate with our standard 80MHz laser, we used an electro-optic pulse picker to select every third pulse, obtaining an effective repetition rate of 27 MHz (38 ns period). Such timing permits up to eight depth layers in principle, although our prototype software currently only handles four layers (see Methods).

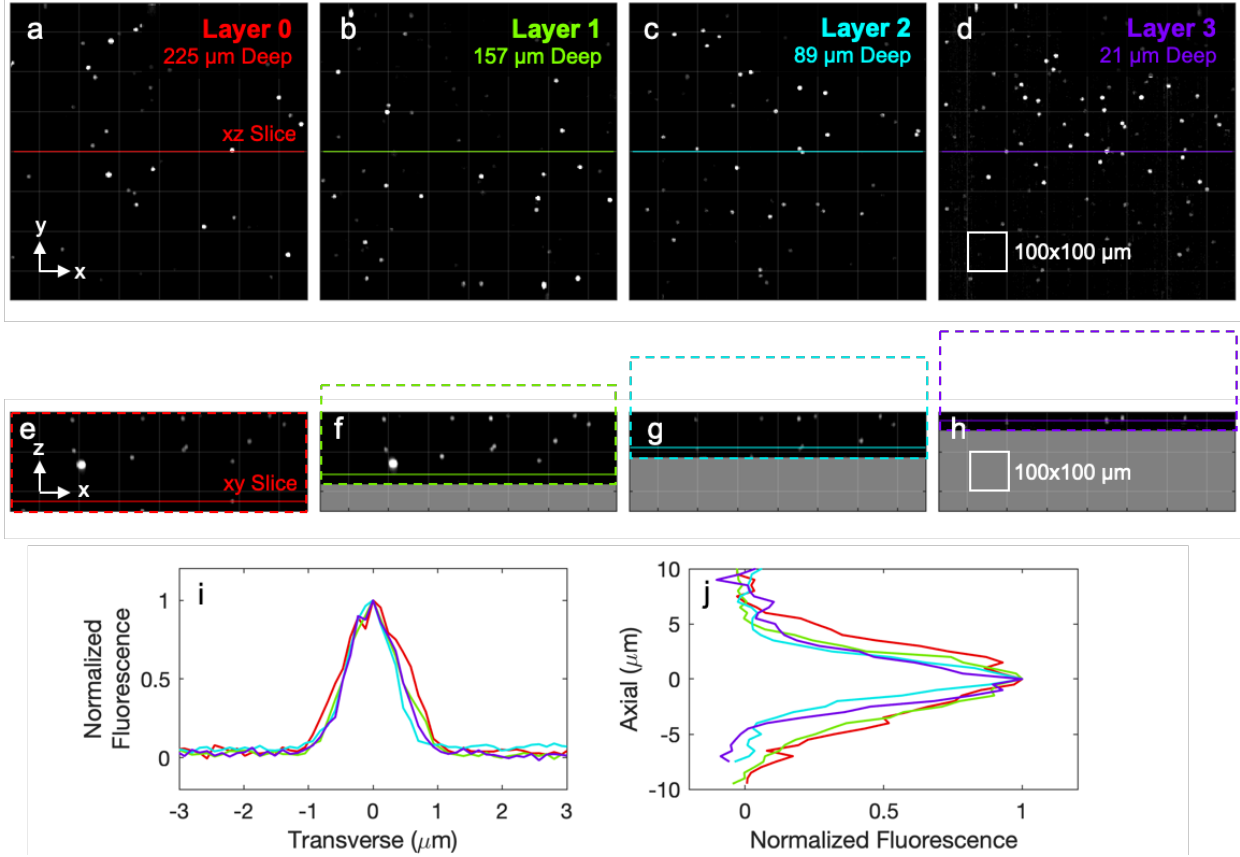

Figure 3: Characterization of reverberation MPM with fluorescent bead samples. **a,b,c,d**, single reverberation image of 10 μm beads in scattering media, with all four layers (68 μm apart) acquired simultaneously, and corrected for crosstalk. **e,f,g,h**, x-z slices obtained from each layer after performing a physical z-scan (dashed boxes), illustrating layer registration. **i,j**, Transverse and axial point spread functions at each layer, as measured with a 1 μm bead.

Initial testing and characterization of our reverberation MPM was done with 10 μm fluorescent beads embedded in a scattering medium ($l_s \approx 100$ μm). A single-shot reverberation image taken at a depth of 225 μm, consisting of four layers spaced 68 μm apart, is shown in the top row of Figure 3. Additionally, Figure 3 shows *x-z* slices obtained from each layer as the sample was vertically scanned by a stage from the surface to 250 μm. The shallower depths, which were separately imaged in different reverberation layers during the extended z-scan, generated the same result with comparable image quality regardless of which layer was used. Profiles of the transverse and axial responses for a single bead are shown in Figure 3, demonstrating that our microscope provides 3D micron-scale resolution similar to a conventional MPM, with a point spread function that is not significantly modified between layers.

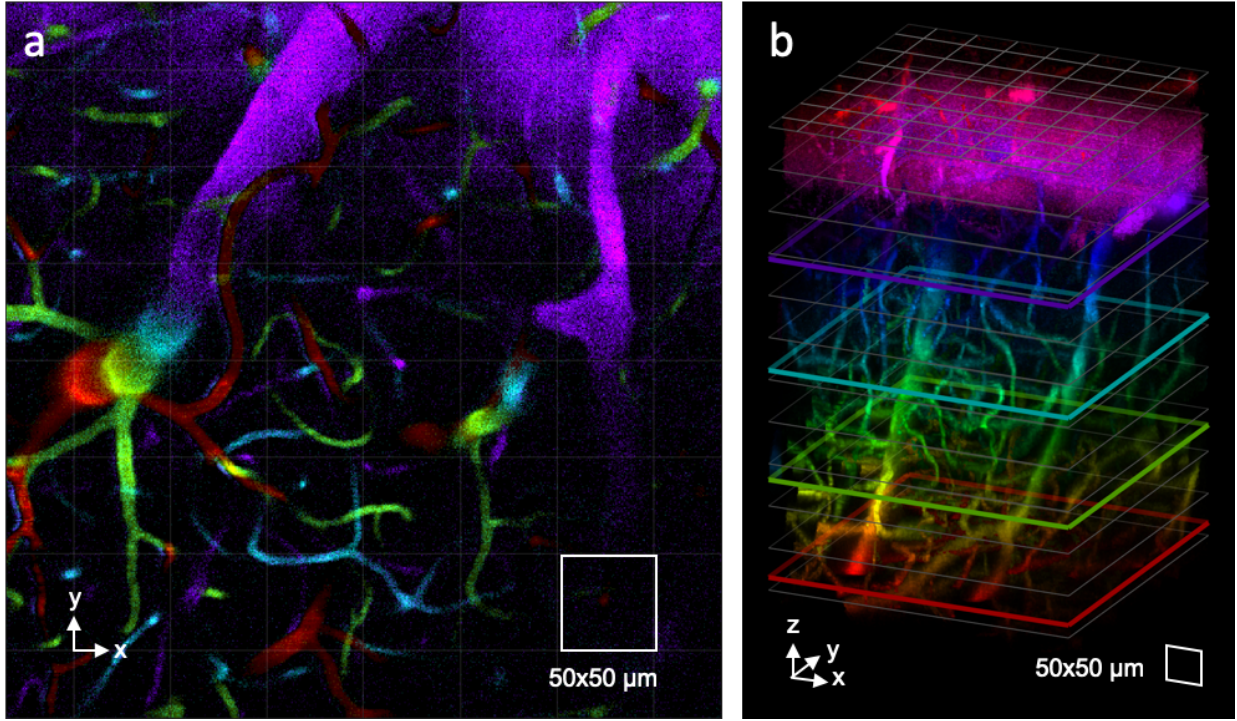

Figure 4: Reverberation imaging of in-vivo mouse-brain vasculature. **a**, single reverberation image comprising four independent layers at different depths; color indicates layer (the single color of each layer merged in overlapping regions). **b**, maximum intensity projection of z-stack with slices from all four reverberation layers merged into a volume; color indicates depth (the z-stack was obtained by a short physical z-scan only to fill in the gaps between reverberation layers – fly-through and fly-around provided in Supplementary Videos 1 and 2). Highlighted layers in **b** correspond to **a**. All images are corrected for crosstalk.

To demonstrate the effectiveness of reverberation MPM for biological imaging, we performed in-vivo imaging of mouse-brain vasculature labeled with FITC-Dextran (Figure 4). In this experiment $l_s$ was found to be approximately 200 μm for an excitation wavelength of 940 nm (in agreement with previous estimates[21]), and the layers were spaced 130 μm apart. The total laser power incident on the sample was 50 mW. These results illustrate our capacity to obtain a comprehensive snapshot of brain tissue over a large depth range, acquired as multiple independent, optically-sectioned layers.

Reverberation MPM presents many advantages for volumetric imaging, with few drawbacks. By splitting each laser pulse into a continuous series of beam foci, multiple layers can be probed near-simultaneously from a large depth (in principle arbitrary) all the way to the sample surface. In cases where depth penetration is laser-power limited, the price paid is a small reduction in the maximum attainable depth penetration by an amount $\Delta z$. In our case, we were limited by an electronic four-channel capacity, allowing us to demonstrate near simultaneous probing of fluorescence over a depth range of ≈400 μm in brain tissue. This range could readily be increased with higher channel capacity, provided the additional layers can fit within the pulse period. Looking forward, our reverberation technique should be of most advantage for ultra-deep MPM imaging with low repetition rate lasers, as used, for example, in three-photon microscopy. Reverberation MPM is both light efficient and simple to implement, requiring only the addition of a reverberation loop to a conventional MPM equipped with fast detection electronics. These advantages make it particularly attractive as a general technique for fast, high resolution, large-scale volumetric imaging in scattering media.

## Methods

**Microscope system:** The laser used was a Coherent Chameleon Ultra II laser (3.5 W tunable Ti:sapphire, 140 fs pulse width) with pulse rate reduced to 27 MHz by a Conoptics pulse selection system (Model 350-210-RA). The laser was typically operated around 940 nm wavelength. Beam steering was performed by Thorlabs GVS001 galvanometer, with z-scanning provided by a Thorlabs MZS500 piezo stage. The objective was a Nikon CFI75

LWD 16× with a numerical aperture of 0.8. Detection was performed by a Hamamatsu H7422PA-40 photomultiplier tube, amplified by a Femto HSA-Y-1-40. Readout was performed by a National Instruments 5771 digitizer and 7972 FPGA combination using customized Vidrio ScanImage software. Note that this software provided only two time bins per channel. We achieved a total of four time bins by exploiting the dual output capacity of our Femto amplifier, and the two-channel capacity our NI digitizer. An Analog Devices AD9516 was used to synchronize the digitizer sampling to the laser pulses. The instrument response time of our detection electronics was confirmed to be better than a nanosecond, as inferred from the signal produced by a second-harmonic crystal sample.

**Mouse preparation:** Mice were anesthetized with ketamine/xylazine, the skin over the dorsal cranium was retracted, and glass imaging windows were implanted over the dorsal neocortex using sterile surgical procedures[24]. Imaging windows and a stainless-steel head post were both anchored with dental acrylic (Metabond, Parkell Inc.). Mice were imaged immediately following surgery. Fluorescent vascular labeling was performed using retroorbital injection of dextran-conjugated FITC (2 MDa, 60 μL of 5% w/v in sterile PBS; Sigma-Aldrich). During imaging sessions, mice were anesthetized with ketamine-xylazine and fixed in a stereotaxic apparatus. All animal procedures were approved by the Boston University Institutional Animal Care and Use Committee and carried out in accordance with NIH standards.

## Data Availability

The data that support the plots within this paper and other findings of this study are available from the corresponding authors upon reasonable request.

## Acknowledgments

We thank Huate Li for initial help in the construction of our reverberation MPM. This work was supported in part by the Engineering Research Centers Program of the National Science Foundation under NSF Cooperative Agreement No. EEC-0812056.

## Author Contributions

J.M. and D.B. conceived of the reverberation technique. D.B. developed and implemented the prototype microscope. I.D. provided the mouse subjects. All authors contributed to experiments, analysis of the data, and the writing of the manuscript.

## Competing Interests

D.B., T.B., and J.M. are coinventors on provisional patent application 62/697,662 submitted by Boston University that covers 'Reverberation Microscopy Systems and Methods'.

## References

1. Helmchen, F. & Denk, W. Deep tissue two-photon microscopy. *Nat. Methods* **2**, 932–940 (2005).
2. Ji, N., Freeman, J. & Smith, S. L. Technologies for imaging neural activity in large volumes. *Nat. Neurosci.* **19**, 1154–1164 (2016).
3. Prevedel, R. *et al.* Fast volumetric calcium imaging across multiple cortical layers using sculpted light. *Nat. Methods* **13**, 1021–1028 (2016).
4. Grewe, B. F., Voigt, F. F., van 't Hoff, M. & Helmchen, F. Fast two-layer two-photon imaging of neuronal cell populations using an electrically tunable lens. *Biomed. Opt. Express* **2**, 2035–46 (2011).

5. Shain, W. J., Vickers, N. A., Goldberg, B. B., Bifano, T. & Mertz, J. Extended depth-of-field microscopy with a high-speed deformable mirror. *Opt. Lett.* **42**, 995 (2017).

6. Sofroniew, N. J., Flickinger, D., King, J. & Svoboda, K. A large field of view two-photon mesoscope with subcellular resolution for in vivo imaging. *eLife* **5**, e14472 (2016).

7. Olivier, N., Mermillod-Blondin, A., Arnold, C. B. & Beaurepaire, E. Two-photon microscopy with simultaneous standard and extended depth of field using a tunable acoustic gradient-index lens. *Opt. Lett.* **34**, 1684 (2009).

8. Kong, L. *et al.* Continuous volumetric imaging via an optical phase-locked ultrasound lens. *Nat. Methods* **12**, 759–762 (2015).

9. Yang, W. *et al.* Simultaneous Multi-plane Imaging of Neural Circuits. *Neuron* **89**, 269–284 (2015).

10. Theriault, G., De Koninck, Y. & McCarthy, N. Extended depth of field microscopy for rapid volumetric two-photon imaging. *Opt Exp* **21**, 10095 (2013).

11. Lu, R. *et al.* Video-rate volumetric functional imaging of the brain at synaptic resolution. *Nat Neurosci* **20**, 620 (2017).

12. Yang, Y. *et al.* Two-Photon Laser Scanning Stereomicroscopy for Fast Volumetric Imaging. *PloS One* **11**, e0168885 (2016).

13. Song, A. *et al.* Volumetric two-photon imaging of neurons using stereoscopy (vTwINS). *Nat Meth* **14**, 420–426 (2017).

14. Amir, W. *et al.* Simultaneous imaging of multiple focal planes using a two-photon scanning microscope. *Opt. Lett.* **32**, 1731–3 (2007).

15. Hu, Q. *et al.* Simultaneous two-plane, two-photon imaging based on spatial multiplexing. *Opt. Lett.* **43**, 4598 (2018).

16. Cheng, A., Gonçalves, J. T., Golshani, P., Arisaka, K. & Portera-Cailliau, C. Simultaneous two-photon calcium imaging at different depths with spatiotemporal multiplexing. *Nat. Methods* **8**, 139–42 (2011).

17. Chen, J. L., Voigt, F. F., Javadzadeh, M., Krueppel, R. & Helmchen, F. Long-range population dynamics of anatomically defined neocortical networks. *eLife* **5**, e14679 (2016).

18. Stirman, J. N., Smith, I. T., Kudenov, M. W. & Smith, S. L. Wide field-of-view, multi-region, two-photon imaging of neuronal activity in the mammalian brain. *Nat Biotechnol* **34**, 857–862 (2016).

19. Heshmat, B., Tancik, M., Satat, G. & Raskar, R. Photography optics in the time dimension. *Nat. Photonics* **12**, 560–566 (2018).

20. Berezin, M. Y. & Achilefu, S. Fluorescence Lifetime Measurements and Biological Imaging. *Chem. Rev.* **110**, 2641–2684 (2010).

21. Theer, P., Hasan, M. T. & Denk, W. Two-photon imaging to a depth of 1000 mm in living brains by use of a Ti:Al2O3 regenerative amplifier. 3 (2003).

22. Beaurepaire, E., Oheim, M. & Mertz, J. Ultra-deep two-photon fluorescence excitation in turbid media. *Opt. Commun.* 5 (2001).

23. Horton, N. G. *et al.* In vivo three-photon microscopy of subcortical structures within an intact mouse brain. *Nat. Photonics* **7**, 205–209 (2013).

24. Holtmaat, A. *et al.* Long-term, high-resolution imaging in the mouse neocortex through a chronic cranial window. *Nat. Protoc.* **4**, 1128–1144 (2009).